\documentclass[aps,prb,superscriptaddress,twocolumn,floatfix,noeprint]{revtex4-2}

\usepackage[utf8]{inputenc}
\usepackage[T1]{fontenc}
\usepackage{mathtools}
\usepackage{dsfont}
\usepackage{bm}
\usepackage{graphicx}
\usepackage[breaklinks]{hyperref}
\usepackage{tikz}
\usepackage{url}
\usepackage{color}
\usepackage{comment}
\usepackage[export]{adjustbox}
\usepackage{float}

\renewcommand{\vec}[1]{\bm{#1}}

\newcommand{\bvec}[1]{\bm{#1}}
\newcommand{\dd}{\mathrm{d}}
\newcommand{\oidx}[4]{o_{#1}o_{#2}o_{#3}o_{#4}} %
\newcommand{\proj}[2]{\mathds P_#1[#2]} %

\DeclareMathOperator{\hadamard}{\circ}

\begin{document}

\allowdisplaybreaks

\title{Functional renormalization group for a large moiré unit cell}

\author{Lennart Klebl} 
\affiliation{Institute for Theory of Statistical Physics,
RWTH Aachen University, and JARA Fundamentals
of Future Information Technology, Germany}
\author{Dante M. Kennes}
\affiliation{Institute for Theory of Statistical Physics,
RWTH Aachen University, and JARA Fundamentals
of Future Information Technology, Germany}
\affiliation{Max Planck Institute for the Structure and Dynamics of Matter, Center for Free Electron Laser Science, 22761 Hamburg, Germany}
\author{Carsten Honerkamp}
\affiliation{Institute for Theoretical Solid State Physics,
RWTH Aachen University, and JARA Fundamentals
of Future Information Technology, Germany}

\date{\today}

\begin{abstract}
  Layers of two-dimensional materials arranged at a twist angle with respect to
  each other lead to enlarged unit cells with potentially strongly altered band
  structures, offering a new arena for novel and engineered many-body ground
  states. For the exploration of these, renormalization group methods are an
  appropriate, flexible tool that takes into account the mutual influence of
  competing tendencies. Here we show that, within reasonable, non-trivial
  approximations, the functional renormalization group known from simpler
  two-dimensional systems can be employed for the large-unit cell moiré
  superlattices with more than 10.000 bands, remedying the need to employ ad hoc
  restrictions to effective low-energy theories of a few bands and/or effective
  continuum theories. This provides a description on the atomic scale, allowing
  one to absorb available ab-initio information on the model parameters and
  therefore lending the analysis a more concrete quantitative character. For the
  case of twisted bilayer graphene models, we explore the leading ordering
  tendencies depending on the band filling and the range of interactions. The
  results indicate a delicate balance between distinct magnetically ordered
  ground states, as well as the occurrence of a charge modulation within the
  moiré unit cell for sufficiently non-local repulsive interaction. 
\end{abstract}

\maketitle
\section{Introduction}

In recent years the field of two-dimensional materials has made major
experimental and theoretical leaps, which led to many fascinating discoveries
and have broadened our spectrum on available phases of matter in these highly
controllable structures. Examples of these novel findings include
superconducting \cite{Xi2015,Benyamini2019} or magnetic \cite{Hurtley966}
phases realized down to the monolayer limit 
and, related to that, the discovery of quantum anomalous hall behavior in thin
films \cite{Chang13,kou2014scale,checkelsky2014trajectory,PhysRevLett.114.187201,%
chang2015high,kandala2015giant,kou2015metal,PhysRevLett.115.126801},
with potentially far-reaching technological applications in the
realm of spin-tronics and quantum computing. 

Recently, the twist angle between two sheets of material stacked atop each other was
added as a further interesting research direction. In these twisted structures
it was shown that, by the emerging huge real space moiré supercells (tiny
Brillouin zone), kinetic energy scales can be reduced drastically, giving rise
to prominent interaction effects. In Refs.~\onlinecite{cao2018unconventional,%
cao2018correlated} it was experimentally
demonstrated that using this route of control two sheets of twisted bilayer
graphene can be tuned to exhibit insulating as well as superconducting behavior.
This exciting finding has spurred an enormous wave of experimental
\cite{yankowitz2018tuning,kerelsky2019maximized,choi2019electronic,Efetov,%
shen2019observation,liu2019spinpolarized,cao2019electric,%
jiang2019charge,wang2019magic,alex2019moirless} as well as theoretical
\cite{doi:10.1021/nl902948m,li2010observation,Bistritzer12233,PhysRevB.86.155449,%
PhysRevLett.109.196802,doi:10.1021/acs.nanolett.6b01906,%
PhysRevX.8.031089,roy2019unconventional,PhysRevX.8.031088,PhysRevX.8.031087,%
Xian2019,procolo2019crucial,TGeSe%
} research. Part of the
excitement is founded in the fact that these two-dimensional systems can be
controlled with relative ease using a backgate, strain or the value of the
twisting angle as a parameter. This allows one to access correlation regimes which
might be otherwise difficult to access in a structure with a chemically
straightforward composition (in this case graphene).
%
%
From a theoretical side a first approach is to concentrate on the
correlation physics if we restrict ourselves to the low-energy bands near the
Fermi surface. However, the validity of  such an approach is difficult to
assess. Often, when considering twisted van der Waals materials (like in the case
of twisted bilayer graphene), there are no band gaps separating the lowest from
higher energy bands and such a separation is unclear. Furthermore,
topological arguments might obstruct the construction of such a simple
low-energy theory \cite{Vishwanath2019Faithful}. In addition, a recent
theoretical study shows that the interplay of correlations might be very subtle
and fragile \cite{throckmorton2019spontaneous}. This calls for a
different vantage point where, in contrast, we do not want to restrict ourselves
to effective low-energy theories neglecting most of the many back-folded bands
arising from the moir\'e supercell. When considering such a theory we immediately
face the problem that at small twist angles many thousand bands need to be kept.
The sheer number of bands makes theoretical descriptions very cumbersome,
especially when one tries to include interaction effects. Recent advances in this direction
include Refs.~\onlinecite{lado2017electrically,xie2018nature,klebl2019inherited}, which treat
correlation effects on the random-phase approximation (RPA) or mean-field level.

Here, we want to add to this by establishing that a more sophisticated tool, the
so-called functional renormalization group (fRG), can be applied to a Hamiltonian
keeping the many bands in the moir\'e Brillouin zone, without reducing to
effective low-energy theories. The fRG is a versatile method capable to describe
a plethora of interacting electron systems
\cite{Metzner2012,Wang2012,Platt2013,Kennes2018,Classen2019}. Yet, solving the full fRG
equations after a truncation at the two-particle vertex ($\Gamma^{(4)}$) is in
general numerically feasible only for systems with a few orbitals per unit cell,
since the vertex function itself scales with the number of orbitals to the
fourth power. In addition, the similarly rich dependence on momenta -- or the unit
cell positions when formulated in real space -- remains a numerical challenge.
Fortunately, for the latter dependence, reasonable simplifications can be
formulated, with justifiable restrictions to short-ranged fermion bilinears
\cite{Husemann2009,Wang2012,Lichtenstein2017,Eckhardt2019,Hille2020}.
Furthermore, at least for a larger set of questions, additional approximations
simplifying the orbital dependence can be made
\cite{honerkamp2018efficient,Honerkamp2018}. Here we employ these two
approximation steps to the fRG equations in order to keep the numerical effort
low enough to treat systems with more than ten thousand orbital sites per unit
cell. We show that the treatment of twisted graphene bilayers close to the
so-called \lq{}magic angle\rq{} using these approximations, resolving all
individual carbon sites in the large moiré unit cell, yields similar results to
what we know from our previous study using the RPA of the
crossed particle-hole channel \cite{klebl2019inherited}.

\begin{figure} 
  \centering
  \vspace{10pt}
  \includegraphics[scale=0.8]{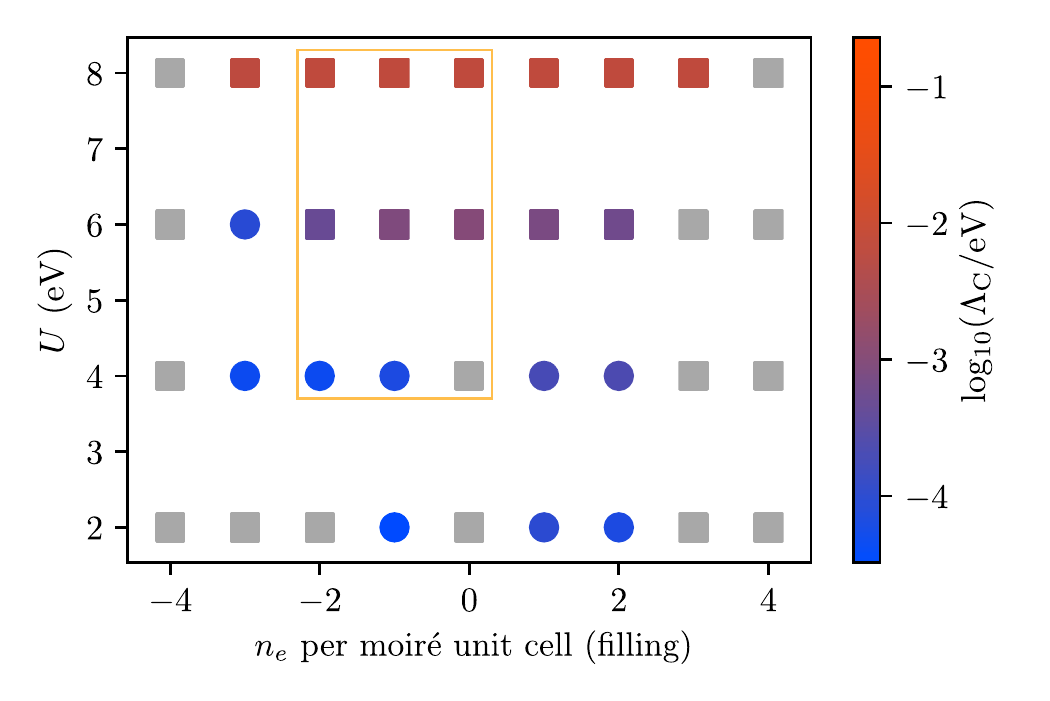}
  \vspace{-16pt}
  \caption{Tentative phase diagram from the $\Gamma$-point IOBI fRG
  for twisted bilayer graphene at angle $1.05^\circ$ using a Hubbard
  interaction. The squares mark nodal antiferromagnetic order, the circles
  ferromagnetic order. For non-diverging flows a gray square is shown. The color
  indicates the critical scale $\Lambda_C$ that roughly corresponds to the
  critical temperature of the phase transition. The orange rectangle shows the
  parameter region which simulations with longer ranged interactions have been
  carried out for (cf.  Fig.~\ref{fig:phaseohno}). At strong interactions, the
  nodal antiferromagnetic phase is favored, whereas at smaller $U$, the system
  is susceptible to ferromagnetic ordering if its filling is close to the van
  Hove singularities of the flat bands.}
  \label{fig:phasehub}
\end{figure}

This work extends such methodology to nonlocal interactions and the coupling to
other channels beyond the RPA, and of course justifies the use of the RPA for
the dominant instability {\itshape a posteriori}. We summarize the main results
of this paper in the tentative phase diagram obtained with this method for a
Hubbard interaction in Fig.~\ref{fig:phasehub}. We find two magnetic orderings:
First, antiferromagnetic order on the atomic scale with a sign change of the
order parameter around the AA regions and, second, ferromagnetic order. The
latter is only found at small interaction strengths and for fillings close to
the van Hove singularities of the material's flat bands at low critical scales.
Nodal antiferromagnetism is present for all fillings that show an instability at
larger couplings.  With increasing nonlocal contributions to the interaction,
the instability gets dominated by the repulsion among electrons outside the AA
regions of the moiré unit cell, indicating an interaction-induced charge
redistribution.  

The rest of this paper is structured as follows: We will shortly introduce the
tight-binding model we use to describe \lq{}magic angle\rq{} twisted bilayer
graphene in Sec.~\ref{sec:model}. Thereafter we discuss the method and
approximations to it in Sec.~\ref{sec:method}. Sec.~\ref{sec:results} shows the
results of our simulations. Finally, we finish with some concluding remarks in
Sec.~\ref{sec:conclusion}.

\section{Model}
\label{sec:model}
\begin{figure} 
  \centering
  \includegraphics[scale=0.8]{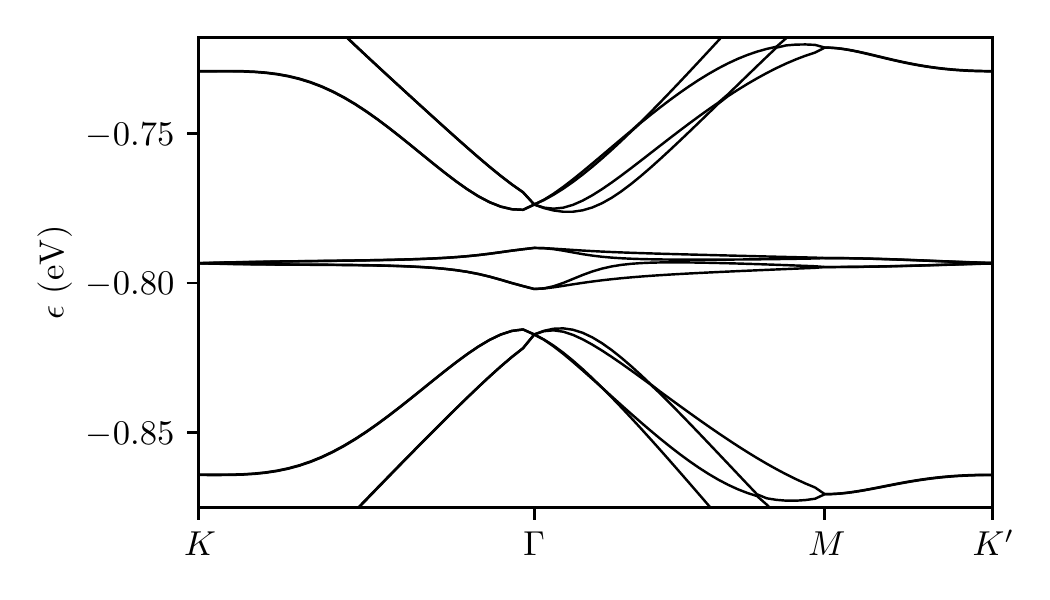} \vspace{-12pt}
  \caption{\label{fig:bands}Low energy window of the non-interacting band
  structure of magic angle twisted bilayer graphene along the irreducible path
  in the moir\'e Brilluoin zone. There are four spin degenerate flat bands
  around charge neutrality at approximately $-0.79\,\mathrm{eV}$.}
\end{figure}

We set up the moir\'e unit cell by constructing the superlattice vectors from 
lattice vectors of the honeycomb lattice of one single graphene sheet, $\bvec
l_1 = (\sqrt3/2,3/2,0)$ and $\bvec l_2 = (\sqrt3,0,0)$. The first superlattice
vector can be written as $\bvec L_1 = n \bvec l_1 + m \bvec l_2$ with $n$ and
$m$ integers defining the twist angle. The second superlattice vector $\bvec
L_2$ is rotated by 60 degrees with respect to $\bvec L_1$. One of the layers is
a honeycomb lattice with Bravais lattice vectors $\bvec l_1$ and $\bvec l_2$,
the other one is shifted vertically and rotated by the twist angle $\theta =
\arccos\frac{m^2+n^2+4mn}{2(m^2+n^2+mn)}$ around the AA site. We implement
corrugation effects by varying the interlayer distance in the supercell as
described in Ref.~\onlinecite{PhysRevX.8.031087}. Choosing $n=31,m=32$ sets
$\theta=1.05^\circ$ and leads to 11908 sites in one supercell. The hopping
parameters are taken from Ref.~\onlinecite{Moon}. Once we have set up the
Hamiltonian for one specific Bloch momentum $\bvec k$ of \lq{}magic angle\rq{}
twisted bilayer graphene, we diagonalize the matrix and obtain the spectrum
$\epsilon_b(\bvec k)$ and the orbital makeup $u_{ob}(\bvec k)$. The low energy
part of the non-interacting band structure shows four (eight including spin)
flat bands around charge neutrality (cf. Fig.~\ref{fig:bands}). From the
spectrum and orbital makeup, we can construct the free (Matsubara) Green's
function as a matrix in orbital space:
\begin{align}
  G_{o_1o_2}^{(0)}(k) = \sum_b \frac{u_{o_1b}^{}(\bvec k) u_{o_2b}^*(\bvec k)}{
    ik_0 - \epsilon_b(\bvec k) + \mu}\,.
\end{align}

As electron-electron interaction, we employ the Ohno ansatz
\cite{Ohno1964,Wehling2011} with a sharp cutoff that allows to treat
increasingly longer ranged interactions. Its functional form in real space reads
\begin{equation}
  V_O(\vec r) = \frac{U\,r_O}{\sqrt{r_O^2+\vec r^2}}\,\Theta(r_c-|\vec r|)
  \ ,
  \label{Vohno}
\end{equation}
where we introduced the Ohno radius $r_O$ and the cutoff radius $r_c$. For our
simulations, we used $r_O=3a$ ($a$ is the graphene lattice constant), $r_c \in
\{0.7a,1.1a,1.5a\}$ (these correspond to nearest, next-nearest and
second-next-nearest-neighbor interactions in mono layer graphene) and the limit
$r_c \rightarrow 0$, i.e. a Hubbard interaction.  Note that there are more
specific descriptions for the non-local interaction, also including the
environmental screening
\cite{Roesner}.

For this qualitative, first fRG study we concentrate on the simple form
Eq.~(\ref{Vohno}) and mainly study the dependence on $r_c$.

\section{Method}
\label{sec:method}
We use the truncated unity approximation to the fRG equations which is described
in more detail in Refs.~\onlinecite{Lichtenstein2017,Hille2020}, building
essentially on earlier works, mainly Refs.~\onlinecite{Husemann2009,Wang2012}.
The main idea of this scheme is to write the interaction vertex as a sum of the
following three channels: direct particle-hole ($D$), crossed particle-hole
($C$) and particle-particle ($P$) channel. Each part can be understood as
interaction between fermion bilinears of the corresponding types. The spatial
structures of these bilinears can then be expanded in basis functions with
specific symmetries in the moiré Brillouin zone.
Even though the basis functions obey specific symmetry relations, they
generally do not restrict the overall symmetry to specific sectors.
This expansion is truncated in
its length.  As a simple but non-trivial approximation we here consider only the
in-cell contribution, i.e. the two fields in the fermion bilinears in pairing,
charge and spin channels are in the same unit cell. Given the large extent of
the moiré unit cell, this already captures quite some spatial dependence and
hence may be a tolerable approximation.  In addition, we then employ a
$\Gamma$-point approximation of the vertex function in the momentum argument of
the three channels.  This means that we ignore all momentum dependence of the
interaction within the small moiré Brillouin zone. One can convince oneself that this provides a
reasonable approximation as long as the interaction decays significantly in real
space through the moiré unit cell. Due to the $\Gamma$-point approximation, we
are sensitive with respect to order parameters that vary spatially within the
moiré unit cell but that do not change when translated into other unit cells.
Again, due to the large unit cell, this does not render the analysis trivial,
but it excludes, e.g., density waves that enlarge the moiré unit cell. These can
be investigated by allowing other wavevectors besides $\Gamma$, which we however
postpone to later work.  In addition, we neglect self-energy contributions as
well as the frequency dependence of the vertex function and truncate the fRG
equations at the four point vertex $\Gamma^{(4)}$. Since we want to treat
$SU(2)$ symmetric systems, we only need to consider the flow equations for the
symmetrized vertex function $V^\Lambda$ where $\Lambda$ is the flow parameter.
The three channels obey coupled differential equations in the flow parameter
$\Lambda$ and carry orbital indices explicitly:
\begin{widetext}
\begin{equation}
  \label{eqn:flow}
  \begin{aligned}
    \frac{\dd}{\dd\Lambda} P^\Lambda_{\oidx{1}{2}{3}{4}} = &{}
  \sum_{o_5\dots o_8} \proj{P}{V^\Lambda_{\oidx{1}{2}{5}{6}}} \dot
  L^{\mathrm{PP},\Lambda}_{o_5\dots o_8}
  \proj{P}{V^\Lambda_{\oidx{7}{8}{3}{4}}} \,,
  \quad
    \frac{\dd}{\dd\Lambda} C^\Lambda_{\oidx{1}{2}{3}{4}} =
  \sum_{o_5\dots o_8} \proj{C}{V^\Lambda_{\oidx{1}{8}{5}{4}}} \dot
  L^{\mathrm{PH},\Lambda}_{o_5\dots o_8}
  \proj{C}{V^\Lambda_{\oidx{6}{2}{3}{7}}} \,, \\
    \frac{\dd}{\dd\Lambda} D^\Lambda_{\oidx{1}{2}{3}{4}} = &{}
    \begin{multlined}[t]
  \sum_{o_5\dots o_8} \Big\{ -2\, \proj{D}{V^\Lambda_{\oidx{1}{8}{3}{5}}} \dot
  L^{\mathrm{PH},\Lambda}_{o_5\dots o_8} \proj{D}{V^\Lambda_{\oidx{2}{6}{4}{7}}}
  + \proj{C}{V^\Lambda_{\oidx{1}{8}{5}{3}}} \dot L^{\mathrm{PH},\Lambda}_{
  o_5\dots o_8} \proj{D}{V^\Lambda_{\oidx{2}{6}{4}{7}}} \\
  +\proj{D}{V^\Lambda_{\oidx{1}{8}{3}{5}}} \dot L^{\mathrm{PH},\Lambda}_{
  o_5\dots o_8} \proj{C}{V^\Lambda_{\oidx{2}{6}{7}{4}}}
  \Big\} \,,
    \end{multlined}
  \end{aligned}
\end{equation}
with the particle-particle and particle-hole loops
\begin{equation}
  \label{eqn:loops}
  L_{\oidx{5}{6}{7}{8}}^{\mathrm{PP},\Lambda} = \frac1{\beta N}\sum_k
  G^\Lambda_{o_5o_7}(k) G^\Lambda_{o_6o_8}(-k) \,, \quad
  L_{\oidx{5}{6}{7}{8}}^{\mathrm{PH},\Lambda} = \frac1{\beta N}\sum_k
  G^\Lambda_{o_5o_6}(k) G^\Lambda_{o_7o_8}(k) \,.
\end{equation}
\end{widetext}
The projection operators $\mathds P$ reduce to unity without further
approximations since they affect the momentum dependencies we disregarded.
In the zero-temperature limit, the Matsubara sums in Eq.~(\ref{eqn:loops})
become integrals. As we only need the differentiated loops
$\dd/\dd\Lambda\,L^{\mathrm{PP/PH},\Lambda}$, it is in some cases possible to
evaluate the integral analytically. For the sharp cutoff we employ in our
simulations, we use the Green's function $G^\Lambda_{o_1o_2}(k) =
G^{(0)}_{o_1o_2}(k)\, \sqrt{\Theta(|k_0|-\Lambda)}$ and are able to trivially
carry out the frequency integrals in the derivative of
$L^{\mathrm{PP/PH},\Lambda}$ via the resulting delta function.

The approximation we make in order to be able to treat systems with a large
number of orbitals is to only allow orbital bilinear interactions in each of the
three interaction channels (IOBI approximation) \cite{honerkamp2018efficient}.
This approximation is certainly valid as long as the interaction mainly consists of density-density type components. On the bare level, this is given as long as the Wannier orbitals used to construct the tight binding model have a negligible overlap. But even beyond this simple situation, the approximation is capable of capturing the standard Kanamori representation of local interactions in terms of intra- and interorbital repulsions and Hund's rule parameters, as described in Ref. \cite{honerkamp2018efficient}.
The IOBI approximation lowers the complexity of each channel of the vertex to
only be number of orbitals to the power of two instead of four. This
simplification reads
\begin{equation}
  \label{eqn:iobi-channels}
  \begin{aligned}
    P^\Lambda_{\oidx{1}{2}{3}{4}} &{}= \delta_{o_1o_2} \delta_{o_3o_4}
    P^\Lambda_{o_1o_3} \,,\\
    C^\Lambda_{\oidx{1}{2}{3}{4}} &{}= \delta_{o_1o_4} \delta_{o_2o_3}
    C^\Lambda_{o_1o_3} \,,\\
    D^\Lambda_{\oidx{1}{2}{3}{4}} &{}= \delta_{o_1o_3} \delta_{o_2o_4}
    D^\Lambda_{o_1o_3} \,.
  \end{aligned}
\end{equation}
The full vertex is composed by simply adding up the three channels:
\begin{align}
  V^\Lambda_{\oidx{1}{2}{3}{4}} &{}= P^\Lambda_{\oidx{1}{2}{3}{4}} +
  C^\Lambda_{\oidx{1}{2}{3}{4}} + D^\Lambda_{\oidx{1}{2}{3}{4}}.
\end{align}
The initial interaction $V^\infty$ is of density-density type and can be written
as a vertex function with components in the $D$ channel only.
$V^\Lambda$ still is a fourth rank tensor in orbital space in the IOBI
approximation, even though each channel is reduced to be a matrix in the orbital
indices. The projections of the full vertex $V^\Lambda$ to a channel are now
simply given by the restrictions of orbital bilinearity:
\begin{equation}
  \begin{aligned}
    \proj{P}{V^\Lambda_{\oidx{1}{2}{3}{4}}} &{}= \delta_{o_1o_2}\delta_{o_3o_4}
    V^\Lambda_{\oidx{1}{2}{3}{4}} \,,\\
    \proj{C}{V^\Lambda_{\oidx{1}{2}{3}{4}}} &{}= \delta_{o_1o_4}\delta_{o_2o_3}
    V^\Lambda_{\oidx{1}{2}{3}{4}} \,,\\
    \proj{D}{V^\Lambda_{\oidx{1}{2}{3}{4}}} &{}= \delta_{o_1o_3}\delta_{o_2o_4}
    V^\Lambda_{\oidx{1}{2}{3}{4}} \,.\\
  \end{aligned}
\end{equation}
Insertion of the channel projections into Equation~(\ref{eqn:flow}) yields the
flow equations in the static IOBI $\Gamma$-point approximation:
\begin{equation}
  \begin{aligned}
    \frac{\dd}{\dd\Lambda} \hat P^\Lambda ={}& \hat V^{\mathds P_P,\Lambda} \,
  \dot{\hat L}^{\mathrm{PP},\Lambda} \, \hat V^{\mathds P_P,\Lambda} \,,\\
    \frac{\dd}{\dd\Lambda} \hat C^\Lambda ={}& \hat V^{\mathds P_C,\Lambda} \,
  \dot{\hat L}^{\mathrm{PH},\Lambda} \, \hat V^{\mathds P_C,\Lambda} \,,\\
    \frac{\dd}{\dd\Lambda} \hat D^\Lambda ={}& -2\,\hat V^{\mathds P_D,\Lambda} \,
  \dot{\hat L}^{\mathrm{PH},\Lambda} \, \hat V^{\mathds P_D,\Lambda} \\
    & +\hat V^{\mathds P_D,\Lambda} \, \dot{\hat L}^{\mathrm{PH},\Lambda} \, \hat
      V^{\mathds P_C,\Lambda}
     +\hat V^{\mathds P_C,\Lambda} \, \dot{\hat L}^{\mathrm{PH},\Lambda} \, \hat
      V^{\mathds P_D,\Lambda} \,.
  \end{aligned}
\end{equation}
The quantities are all matrices and connected by matrix products. The projected
vertex function's matrices read
\begin{equation}
  \begin{aligned}
    \hat V^{\mathds P_P,\Lambda} &{}= \hat P^\Lambda + \mathrm{diag}(\hat
    C^\Lambda) + \mathrm{diag}(\hat D^\Lambda) \,,\\
    \hat V^{\mathds P_C,\Lambda} &{}= \mathrm{diag}(\hat P^\Lambda) + \hat
    C^\Lambda + \mathrm{diag}(\hat D^\Lambda) \,,\\
    \hat V^{\mathds P_D,\Lambda} &{}= \mathrm{diag}(\hat P^\Lambda) +
    \mathrm{diag}(\hat C^\Lambda) + \hat D^\Lambda \,.
  \end{aligned}
\end{equation}
By the IOBI approximation, the loops are constrained to be matrices as well and
take the following form:
\begin{equation}
  \begin{aligned}
    \hat L^{\mathrm{PP},\Lambda} &{}= \int\frac{\dd k_0}{2\pi}\, \frac1N\sum_{\vec k} \hat G^\Lambda(k)
    \hadamard \hat G^\Lambda(-k) \,,\\
    \hat L^{\mathrm{PH},\Lambda} &{}= \int\frac{\dd k_0}{2\pi}\, \frac1N\sum_{\vec k} \hat G^\Lambda(k)
    \hadamard \hat G^\Lambda(k) \,,
  \end{aligned}
\end{equation}
where the operation $\hat A \hadamard \hat B$ is an element-wise matrix product
of the matrices $\hat A$ and $\hat B$.

\begin{figure}
    \centering
    \includegraphics[scale=0.9]{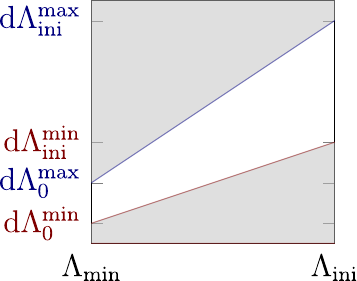}
    \caption{Sketch of envelope functions to choose $\mathrm{d}\Lambda$ as a
    function of $\Lambda$ for the integration of the fRG flow. The gray area is
    inaccessible, $\mathrm{d}\Lambda$ has to be between the red and blue
    functions in the white area. In our simulations, we restricted
    $\mathrm{d}\Lambda$ by two power-law boundaries with exponent $1.1$ in order
    to maximize numerical efficiency.}
    \label{fig:lambda-bdy}
\end{figure}
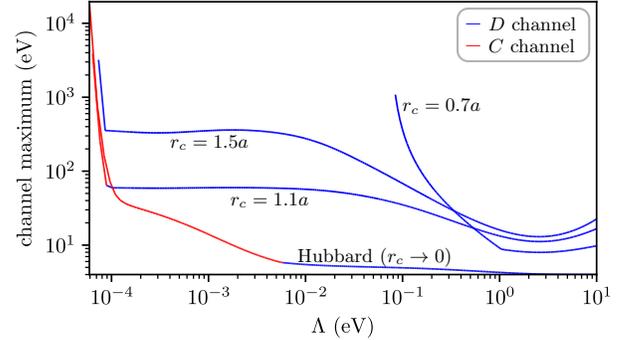
\begin{figure} 
  \centering
  \begin{tikzpicture}
    \def\scx#1{\scalebox{0.8}{#1}}
    \node[anchor=south west] at (0,0) {%
      \includegraphics[scale=0.8]{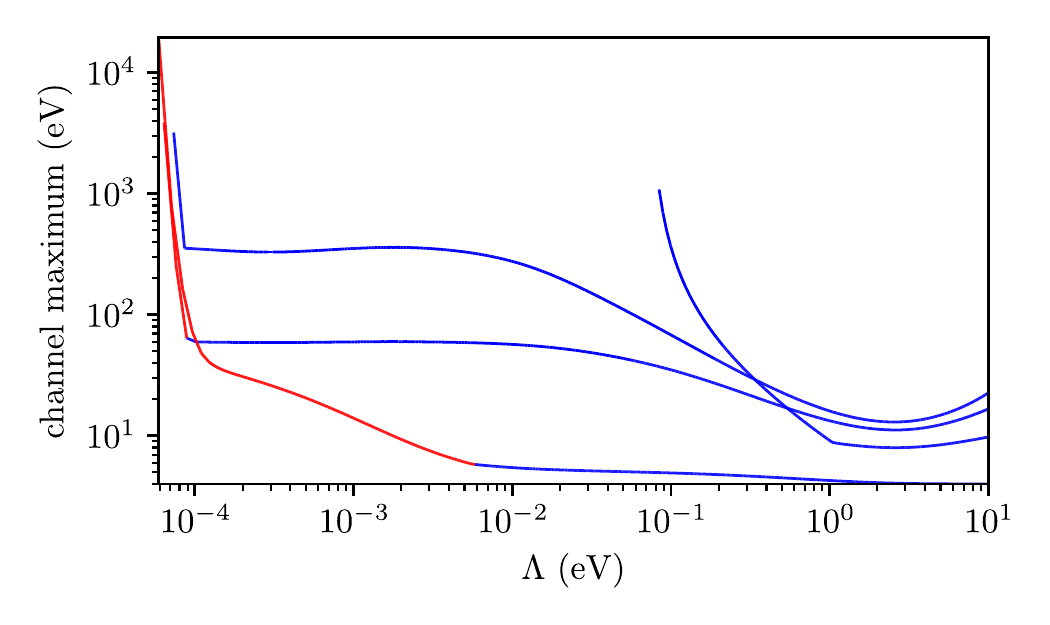}
    };
    \node[anchor=south] at (5.2,1.25) {\scx{Hubbard ($r_c\rightarrow0$)}};
    \node[anchor=west] at (5.45,3.5) {\scx{$r_c=0.7a$}};
    \node[anchor=north] at (3.8,2.46) {\scx{$r_c=1.1a$}};
    \node[anchor=north] at (3,3.23) {\scx{$r_c=1.5a$}};
    \draw[rounded corners,color=gray!60,thick] (6.3,4.1) rectangle (8.05,4.8) {};
    \node[anchor=north west] at (6.3,4.8) {{\bf\color{blue}--}\ \scx{$D$ channel}};
    \node[anchor=south west] at (6.3,4.1) {{\bf\color{red}--}\ \scx{$C$ channel}};
  \end{tikzpicture}
  \vspace{-14pt}
  \caption{\label{fig:flows}fRG flow of the maximal channel of the interactions
  with different interaction cutoffs $r_c$ and onsite Hubbard interaction
  ($r_c\rightarrow0$). The lines' colors represent whether the vertex function's
  maximum occurs in the $C$ channel (red) or the $D$ channel (blue).  The system
  is doped to $-1$ electron per moir\'e unit cell with an onsite interaction
  strength $U=4\,\mathrm{eV}$. For a cutoff at nearest-neighbor interaction
  ($r_c=0.7$), a $D$ channel instability is favored whereas for the other
  interactions, it has a $C$ channel instability at much lower critical value
  $\Lambda_C$.}
\end{figure}
\begin{figure*}
  \centering
  \makebox[12pt]{(a)}
  \includegraphics[scale=0.9,valign=t]{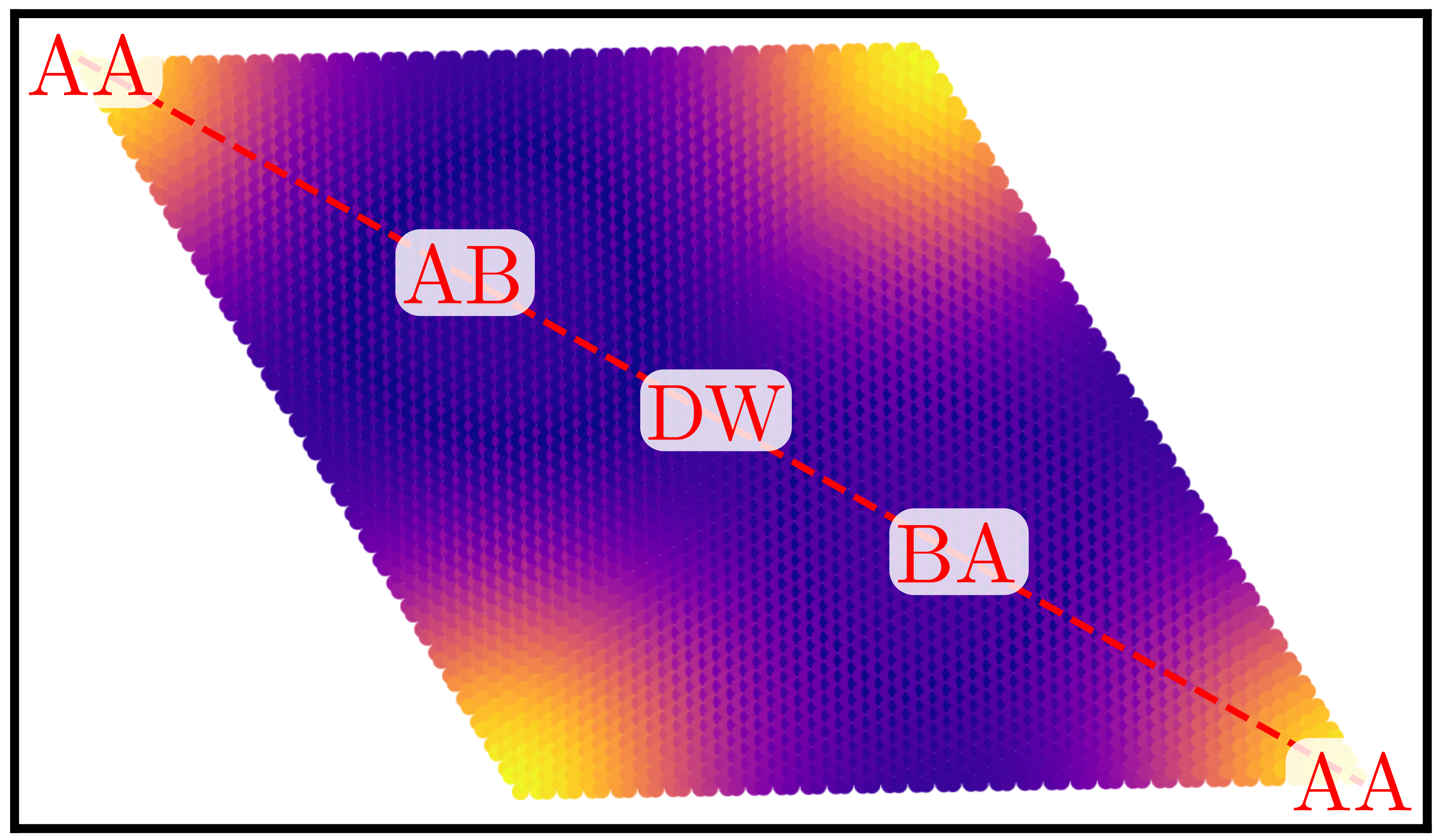}
  \includegraphics[scale=0.9,valign=t]{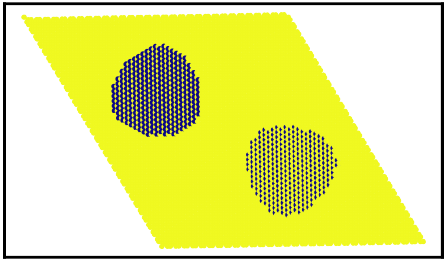}
  \includegraphics[scale=0.8,valign=t]{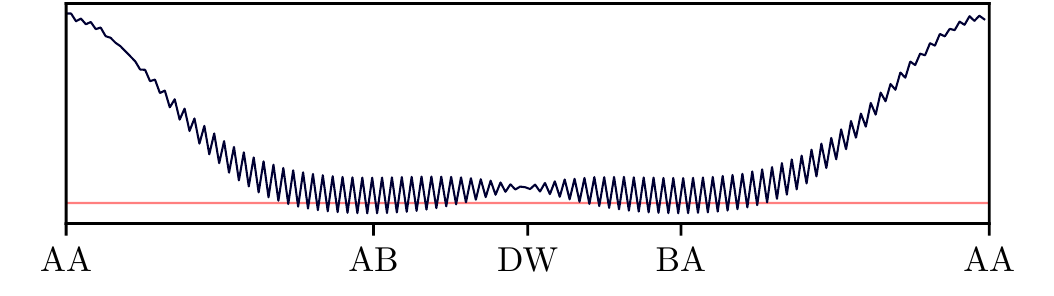}
  \\
  \makebox[12pt]{(b)}
  \includegraphics[scale=0.9,valign=t]{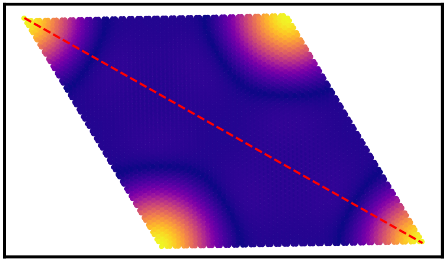}
  \includegraphics[scale=0.9,valign=t]{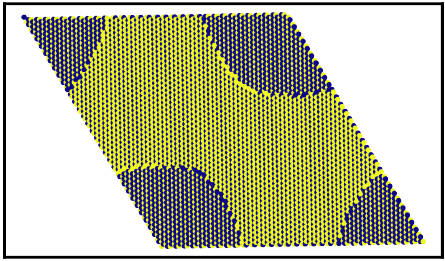}
  \includegraphics[scale=0.8,valign=t]{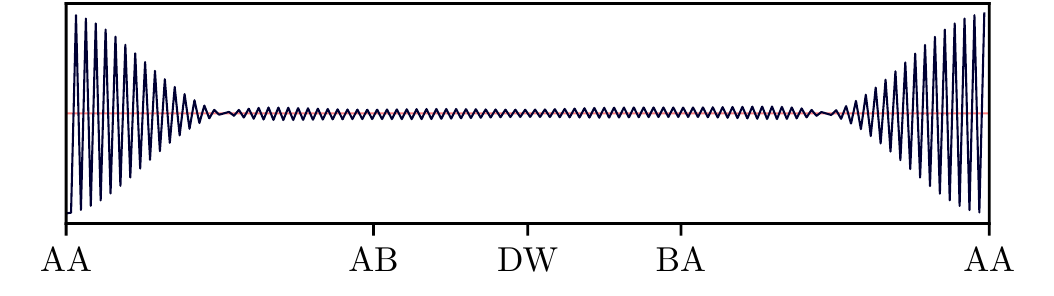}
  \\
  \makebox[12pt]{(c)}
  \includegraphics[scale=0.9,valign=t]{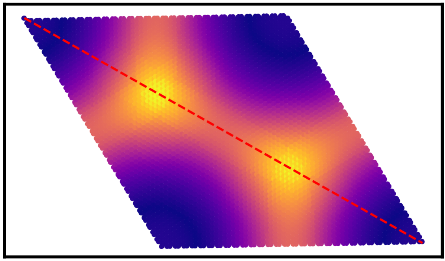}
  \includegraphics[scale=0.9,valign=t]{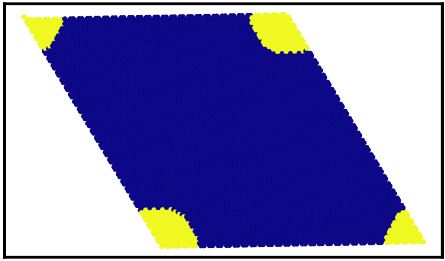}
  \includegraphics[scale=0.8,valign=t]{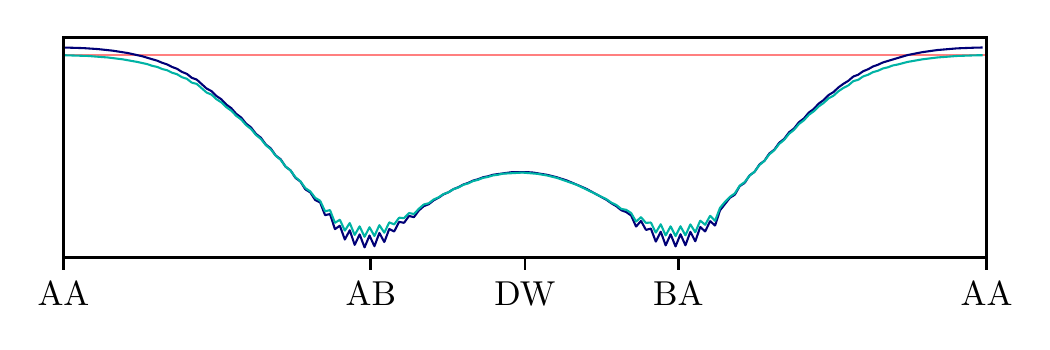}
  \caption{The three different orderings found in this work represented by the
  eigenvectors of the leading channel at the end of the fRG flow. (a):
  ferromagnetic ordering, $C$ channel, Hubbard interaction ($r_c\rightarrow0$),
  $-1$ electron per moir\'e unit cell, $U=4\,\mathrm{eV}$, (b): nodal
  antiferromagnetic ordering, $C$ channel, Hubbard interaction
  ($r_c\rightarrow0$), charge neutrality, $U=6\,\mathrm{eV}$, (c):
  charge-modulation ordering eigenvector, $D$ channel, Ohno interaction with
  $r_c=1.5a$, $-1$ electron per moir\'e unit cell, $U=6\,\mathrm{eV}$. The left
  column is the absolute value of the corresponding eigenvector for the top
  layer (blue corresponds to zero, yellow to the maximum), the center column is
  the sign (blue: minus, yellow: plus) and the right column is the value along
  the dashed red line through the rhomb shaped unit cell. In the right panel of
  (c), the cyan line indicates another realization of $D$ channel
  charge-modulation ordering where no sign change in the eigenvector is present.
  It corresponds to the onsite interaction strength $U=8\,\mathrm{eV}$.}
  \label{fig:allorderings}
\end{figure*}
Starting from a value $\Lambda_\mathrm{ini} = 10\,\mathrm{eV}$ for the frequency
cutoff, we integrate the flow to $\Lambda_0 = 0$. The value of the
frequency step $\mathrm{d}\Lambda$ as a function of $\Lambda$ is adaptively
chosen in the window given by two envelope functions as shown in
Fig.~\ref{fig:lambda-bdy}.
Quite generally, the fRG flow leads to strong coupling, i.e. a rapid growth of
certain components of the flowing interaction. We draw physical conclusions from
this by calculating the maximum eigenvalue of each interaction channel for each
step for the intraorbital bilinear
matrices $\hat P^\Lambda, \hat C^\Lambda$ and $\hat D^\Lambda$. If one of the
eigenvalues surpasses a critical value (which we set to
$10^3\,\mathrm{eV}$), we stop the flow and do a full eigendecomposition of the three
channel matrices. The eigenvector corresponding to the maximum eigenvalue
indicates the orbital character of the order parameter associated to the
divergence. An example of several fRG flows at fixed filling and interaction
strength for different interaction cutoffs is shown in Fig.~\ref{fig:flows}.
Including nearest-neighbor interaction leads to a divergence in the
$D$ channel compared to the magnetic instability in the $C$ channel found for
onsite and longer ranged interactions.

\section{Results}
\label{sec:results}
The three main types of ordering found as run-away flows in this work are shown
in terms of their leading eigenvectors in Fig.~\ref{fig:allorderings}: (a)
ferromagnetic order throughout the unit cell with some residual
antiferromagnetism on the carbon-carbon-bond scale in the AB regions, (b) nodal
antiferromagnetic order with a sign change of the antiferromagnetic order
parameter on the atomic scale around the AA regions and (c) charge-modulated
states that form a honeycomb lattice of the AB and BA regions. This last
instability is characterized by a leading eigenvector in the density-channel. It
should be analyzed in its pure form for $r_c=1.5a$ and $U=8\,\mathrm{eV}$ where
the sign change in the eigenvector occurring for smaller $r_c$ and $U$ has
disappeared. Then the charge modulation arises from the leading eigenvector that
has largest weight in the AB-regions. This represents a strong growth of the
effective electronic repulsion in those regions, which should then push the
charges away from there into the AA-regions.

\subsection{Hubbard interaction ($r_c\rightarrow0$)}
For onsite interaction, we find instabilities of magnetic type
(cf.~Fig.~\ref{fig:allorderings},~(a),~(b)) that agree with
the types of ordering we found in our previous RPA study for the magnetic
susceptibility due to a pure onsite interaction \cite{klebl2019inherited}.
Additionally, the values of $\Lambda_C$ are similar in their order of magnitude
to the critical temperatures found in RPA. The phase diagram obtained from fRG
using a Hubbard interaction ($r_c\rightarrow0$) is shown in
Fig.~\ref{fig:phasehub}.

Ferromagnetism is present for fillings close to the van Hove singularities of
the flat bands and small interaction strengths. Increasing $U$ or doping away
from the van Hove singularities leads to nodal antiferromagnetism.

\begin{figure*}
  \centering
  \makebox[12pt]{(a)}
  \includegraphics[scale=0.9,valign=t]{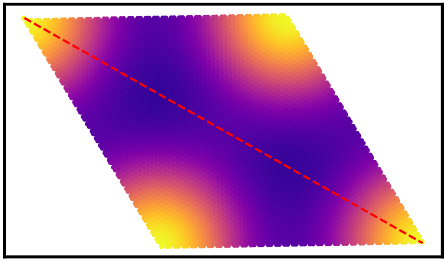}
  \includegraphics[scale=0.9,valign=t]{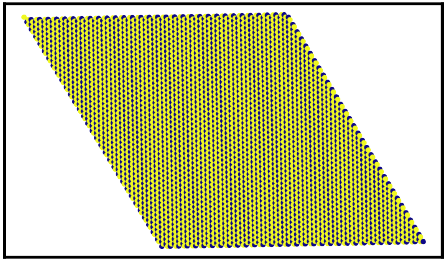}
  \includegraphics[scale=0.8,valign=t]{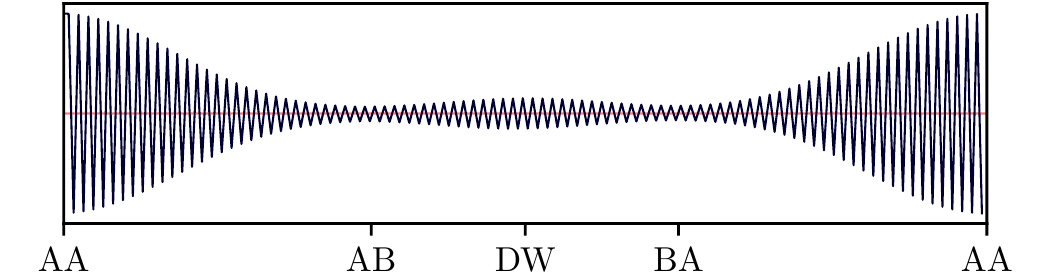}
  \hfill\null
  \\
  \makebox[12pt]{(b)}
  \includegraphics[scale=0.9,valign=t]{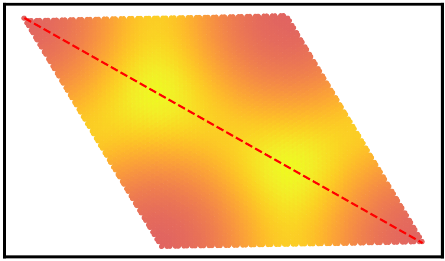}
  \includegraphics[scale=0.9,valign=t]{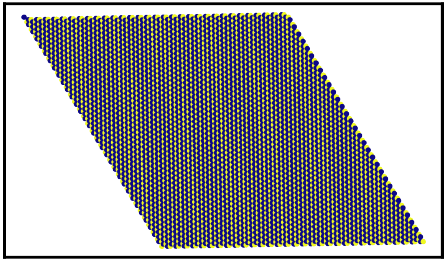}
  \includegraphics[scale=0.8,valign=t]{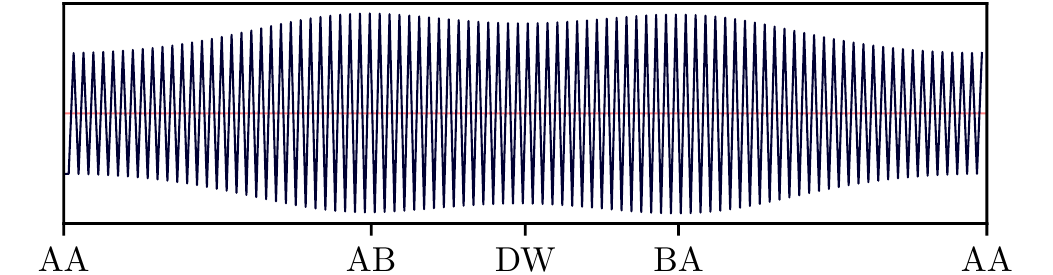}
  \hfill\null
  \\
  \makebox[12pt]{(c1)}
  \includegraphics[scale=1.0,valign=t]{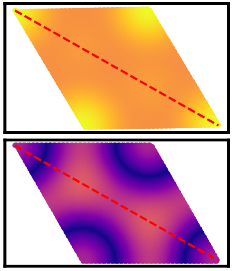}
  \includegraphics[scale=1.0,valign=t]{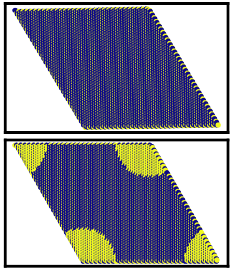}
  \hfill
  \hspace{0.2em}
  (c2)
  \includegraphics[scale=1.0,valign=t]{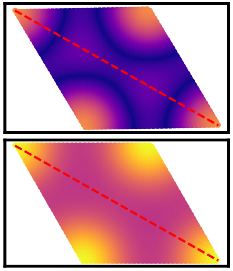}
  \includegraphics[scale=1.0,valign=t]{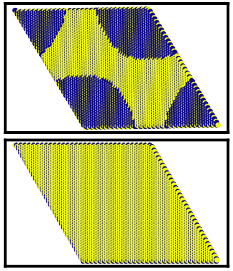}
  \hfill\null
  \\
  \caption{Charge density wave (CDW) orderings found for a cutoff at
  nearest-neighbor interaction ($r_c=0.7a$). (a): $U=4\,\mathrm{eV}$, (b):
  $U=8\,\mathrm{eV}$ and (c1),(c2): $U=6\,\mathrm{eV}$. The leading
  eigenvector of the $D$ channel is shown. As in Fig.~\ref{fig:allorderings},
  the left column is the absolute value of the eigenvector for the top
  layer (blue corresponds to zero, yellow to the maximum), the center column the
  sign (blue: minus, yellow: plus) and the right column the value along the
  dashed red line through the rhomb shaped unit cell. (c1) shows the leading and
  (c2) the subleading eigenvector for both layers. Note that the sign change of
  the staggering around the AA sites is in the lower layer for the leading
  eigenvector and in the upper layer for the subleading eigenvector.}
  \label{fig:special-d-chan}
\end{figure*}
\subsection{Longer ranged interactions}
\begin{figure}
  \centering
  \includegraphics[scale=0.8]{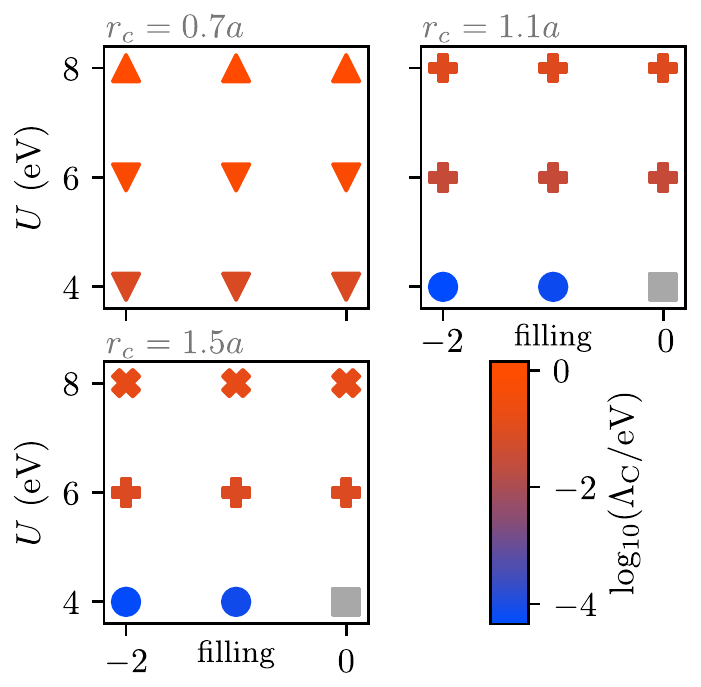}
  \caption{Dependence of the phase diagram of magic angle
  twisted bilayer graphene on the range of the interaction $r_c$.  Squares mark
  nodal antiferromagnetic order, circles ferromagnetic order, plus symbols CM
  states without sign change and crosses CM states with sign change. For
  $r_c=0.7a$, a CDW state (on the carbon-carbon-bond scale) with slight
  amplitude modulation on the moir\'e scale is found. These states are marked by
  triangles.  Gray squares indicate that the vertex did not diverge before
  reaching $\Lambda = 0$.}
  \label{fig:phaseohno}
\end{figure}

\begin{figure}
    \centering
    \includegraphics[width=0.95\columnwidth]{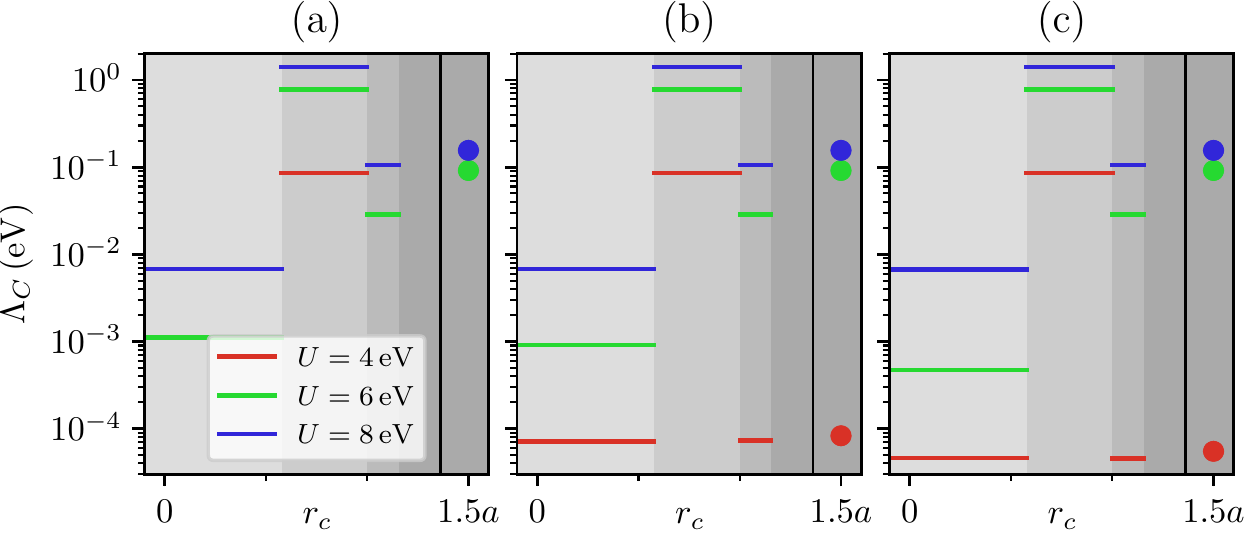}
    \caption{Dependence of $\Lambda_C$ on range of the interaction $r_c$. (a)
    shows the results for charge neutrality, (b) for doping of $-1$ electron per
    moir\'e unit cell and (c) for $-2$ electrons per moir\'e unit cell. The CM
    ordering at $U>4\,\mathrm{eV}$ and $r_c \ge a$ consistently shows an increase
    of $\Lambda_C$ for all fillings. The differently shaded regions indicate the
    regions where the interaction includes the first, second, third and fourth
    nearest neighbor. Beyond the black vertical line, interlayer couplings
    are present in the initial interaction.}
    \label{fig:lambda-mu}
\end{figure}

For longer ranged interactions, we find that the system is susceptible to a
charge modulation (CM) or charge density wave (CDW) instability. For the
fillings and interaction strengths indicated by the orange square in
Fig.~\ref{fig:phasehub}, we carried out the simulations with longer range
interactions. The results are summarized in a series of additional tentative
phase diagrams in Fig.~\ref{fig:phaseohno}. We observe that for the cases where
magnetism is found ($r_c>0.7a$, $U=4\,\mathrm{eV}$), the critical scale
$\Lambda_C$ is almost independent of whether the interaction is longer ranged or
of Hubbard type ($r_c\rightarrow0$).  CM ordering becomes relevant for larger
$U$ ($r_c>0.7$).  While the quantitative location of the AF-to-CM
transition has to be checked in more elaborate numerical studies, our current
finding indicates a high sensitivity of the interacting system with respect to
non-local interaction, as also stated in
Ref.~\onlinecite{throckmorton2019spontaneous}. Additionally, the critical scale
takes much larger values for both the CM and CDW instabilities (see
Fig.~\ref{fig:lambda-mu}).

Including nearest-neighbor interactions $(r_c=0.7a$) leads to charge
density wave (CDW) states as dominant instability. In the context of inheriting
the instabilities from AA and AB stacked graphene bilayers, charge density wave
states with opposite charge density modulation on A and B carbon sublattices are
expected at longer ranged interactions \cite{delapena2014,SchererBi}.
Fig.~\ref{fig:special-d-chan}
shows the CDW states on the carbon-carbon-bond scale with slight modulations on
the moir\'e scale. The case $U=6\,\mathrm{eV}$ shows two almost degenerate
orderings with (approximate) layer degeneracy.

As soon as next-nearest-neighbor interactions are included, $(r_c > 0.7$) the
CDW states for $U\ge6\,\mathrm{eV}$ get replaced by CM states that feature a
variation of the effective density-density interaction on the much longer moiré scale and no staggering on the carbon-carbon-bond scale (cf.~Fig.~\ref{fig:allorderings},~(c)). This interaction is expected to drive a charge transfer within the unit cell. As the interaction is stronger in the AB regions, as visible from the absolute magnitude of the eigenvector in Fig.~\ref{fig:allorderings},~(c), the charge density should get lowered there. 

To support the validity of the fRG method used in this study, we
set up simulations of single layer graphene in a supercell of approximately one
thousand sites and could reproduce the results for varying nearest- and
next-nearest-neighbor interaction strengths found in fRG studies with high
wavevector resolution \cite{pena2017competing} (not shown). This earlier study also showed a variation of the CDW wavevector toward the $\Gamma$-point for increasingly nonlocal interactions, consistent with the findings in the moiré systems here.  


\section{Conclusion}
\label{sec:conclusion}
We use the functional renormalization group to describe interactions and their
ordering tendencies in large moiré unit cells. The results for twisted bilayer
graphene at magic angle with pure onsite interactions agree with our previous
RPA study.  Using the fRG allows us to include long-range interactions in
addition.  Here, we find charge modulated and charge density wave states besides
magnetic orderings and investigate the competition between the different
ordering tendencies.

Until now, only $s$-wave correlations are allowed by our approximations. Even in
this approximation scheme, we are able to show how different order parameters
emerge, some of which are inherited from non-twisted graphene bilayer systems
(antiferromagnetism on the atomic scale and charge density wave on the atomic
scale).  Additionally, the longer ranged Ohno interaction that is accessible in
fRG leads to a competition between direct- and crossed particle hole channel,
i.e.,  charge and magnetic fluctuations. As a strong competitor of the magnetic
order we find a intra-moiré-cell charge-modulated state for which the Coulomb
repulsion in the AB-regions between the AA-spots flows to strong coupling. To
the best of our knowledge, this instability has not been discussed before and
deserves further studies. In particular one should understand whether this state
can help explain the phenomenology of twisted-bilayer graphene systems, e.g. by
depleting the low-energy density of states between the AA regions even further
and therefore localizing electrons in nearly isolated AA islands. Furthermore,
it might well be that, once the charge redistribution has been accounted for,
e.g. by self-energy terms, the previous magnetic instabilities become relevant
again beyond the parameter range found in this work.

The main goal of this work was to demonstrate that one can set up functional
renormalization group calculations within  useful approximations even for such
systems with $\mathcal{O}(10^4)$ bands.  We plan to extend the method to allow
for more quantitative comparisons and also to treat other pairing bilinears in
the $P$ channel and momentum dependencies in the other channels. This will allow
us to study, e.g.,  unconventional superconductivity and other bond ordering
phenomena on the moir\'e scale in the IOBI fRG for systems like \lq{}magic
angle\rq{} twisted bilayer graphene.

\section{Acknowledgements}
The German Science Foundation (DFG) is acknowledged for support through RTG
1995 and under Germany's Excellence Strategy - Cluster  of  Excellence  Matter  and  Light  for Quantum Computing (ML4Q) EXC 2004/1 - 390534769. Support by the Max Planck Institute - New York City  Center for Non-Equilibrium Quantum Phenomena is acknowledged. Most simulations were performed with computing resources granted by RWTH
Aachen University under project rwth0496. In addition, the authors also
gratefully acknowledge the computing time granted through JARA on the
supercomputer JURECA at Forschungszentrum J\"ulich \cite{jureca}.

\bibliography{Bib_moire}

\end{document}